\documentstyle[sprocl,epsfig]{article}

\def\NPB{{\em Nucl. Phys.} B}

\def\PRD{{\em Phys. Rev.} D}

\def\be{\begin{equation}}
\def\ee{\end{equation}}
\def\bea{\begin{eqnarray}}
\def\eea{\end{eqnarray}}

\begin{document}

\hfill
\begin{minipage}[t]{3in}
\begin{flushright}
UCB--PTH--00/01\\
January 2000\\
\end{flushright}
\end{minipage}
\vskip 1cm

\title{FINAL STATE INTERACTION IN HEAVY HADRON DECAY}

\author{MAHIKO SUZUKI}

\address{Department of Physics and Lawrence Berkeley National Laboratory\\
University of California, Berkeley, CA 94720, USA\\E-mail: msuzuki@lbl.gov} 




\maketitle\abstracts{I present a critical account of 
the final-state interaction (FSI) in two-body $B$ decays from viewpoint of 
the hadron picture. I emphasize that the phase and the magnitude of 
decay amplitude are related to each other by a dispersion relation. In 
a model phase of FSI motivated by experiment, I illustrate how much the 
magnitude of amplitude can deviate from its factorization value by the FSI.}

\section{Does the FSI phase diminish with the initial hadron mass ?}
     For a limited number of decay modes, we can extract the FSI phase 
directly from experiment. It is large for the $D$ decays \cite{Bis} and 
small for the $B$ decays: \cite{MS1} 
\begin{eqnarray}
       \Delta\delta &=& 80^{\circ}\pm 7^{\circ}; \;\; D\rightarrow
                                         K^-\pi^+/\overline{K}^0\pi^+, \\
       \Delta\delta &<& \left\{ \begin{array}{ll}
                          11^{\circ};& B\rightarrow
                                         D^+\pi^-/D^0\pi^-\\ 
                          16^{\circ};& B\rightarrow
                                         D^+\rho^-/D^0\rho^-\\
                          29^{\circ};& B\rightarrow
                                         D^{*+}\pi^-/D^{*0}\pi^- .\\
                        \end{array} \right.
\end{eqnarray}
Though nothing is known for $B\rightarrow \pi\pi, K\pi, K\overline{K}$,
we might conjecture that the FSI phases diminish as the initial mass 
increases. Meanwhile the relative FSI phase is known for
the $ggg$ and ''$\gamma$'' decay amplitudes of the 
$J/\psi$ decays: \cite{MS2}
          $\Delta\delta \approx 90^{\circ}$ and $75^{\circ}$
for $J/\psi\rightarrow 0^-0^-$ and $0^-1^-$, respectively.
Therefore, if the FSI phases diminish with the increasing initial mass,
it should start diminishing somewhere
between $m_{J/\psi}$ and $m_B$ or at higher energies.

Can we argue that short-distance (SD) QCD should 
dominate \cite{BJ} in the FSI of $B$ meson decays ? 
I am skeptical. Take the spectator decay $B^+(\overline{b}u)\rightarrow 
K^0(\overline{s}d)\pi^+(\overline{d}u)$.
$K^0$ is formed with energetic $\overline{s}$ and $d$
flying away colinearly, while $\overline{d}$ picks up the 
soft spectator $u$ to form $\pi^+$. 
The gluons exchanged between $\overline{d}$ and $\overline{s}$ 
(or $d$) are hard. But the gluons between the soft spectator $u$ and 
$\overline{s}$
(or $d$) are not so hard. The CM energy of the $u$ and $\overline{s}$ 
(or $d$) grows only with $\sqrt{m_b}$, not linearly in $m_b$. 
By simple kinematics,  
$m_{\overline{d}u} \approx (\Lambda_{\rm QCD}m_b)^{1/2}\simeq$1.2 GeV 
for $E_u\simeq
\Lambda_{\rm QCD}$, which is in the middle of the resonance region of the
${\overline{s}d}$ channel. Then long-distance (LD) interactions cannot be 
ignored between $K^0$ and $\pi^+$. One may stick to the quark picture
by invoking the local quark-hadron duality.\cite {SD} 
While it may be appropriate for inclusive decays, the 
duality is a dynamical hypothesis yet to be tested in two-body B decays.

\section{Why is the FSI complicated in the $B$ decay ?}
   The FSI phases of $K\rightarrow \pi\pi$  and the hyperon decay
are equal to the strong interaction phase shifts by the phase theorem,
since the final states are the eigenstates of {\em S} matrix. In contrast,
the FSI phases of the $B$ decay depend on both strong and weak interactions.
Take the decay $B\rightarrow K\pi$.  Since $K\pi$ 
is made of a large number of eigenchannels $a$ of {\em S} matrix,
        $|K\pi^{out}\rangle = \sum_a c_a|a^{out}\rangle$,
the decay amplitude through the operator $\cal O$$_i$ 
acquires an eigenphase $\delta_a$ from each eigenchannel as
\begin{equation}
        M^{(i)}(B\rightarrow K\pi) = \sum_a c_a M^{(i)}_{a}e^{i\delta_a},
                    \;\;(c_a^* = c_a; M^{(i)*}_a = M^{(i)}_a). \label{1}
\end{equation} 
The net phase of $M^{(i)}$ depends on ${\cal O}_i$ even 
for the same isospin eigenstate of $K\pi$.  
The FSI phase has nothing to do with the phase shift of elastic 
$K\pi$ partial-wave amplitude,
$\eta e^{i\delta_{K\pi}}\sin\delta_{K\pi}
\equiv(\sum_a|c_a|^2 e^{2i\delta_a}-1)/2i$ at the $B$ mass. 

\section{Meson-meson interaction at $E_{\rm cm}\simeq$ 5 GeV}

To understand the FSI, we need to know how two mesons 
interact at $E_{\rm cm} = m_B$. The elastic meson-meson 
scattering amplitude can be written as
\begin{equation}
 T(s,t) =  i\sigma_{\rm tot}se^{bt}+\sum_i\gamma_i(t)s^{\alpha_i(t)} + \cdots. 
                                       \label{inv}
\end{equation}
The dominant term is the Pomeron exchange, which is purely imaginary and 
sharply peaked in the forward direction. With a bit of theory,\cite{SW}
we find $\sigma_{\rm tot}= 15\sim 20$mb and $b\approx 2.8 {\rm GeV}^{-2}$
for $K\pi$. These values are much the same for other meson 
scatterings. The non-Pomeron terms are 
down by $\sqrt{s}$ at $t=0$.

The partial-wave projection of $T(s,t)$ gives $a_0(m_B)\approx 0.17i$.
It leads to a large inelasticity:
\begin{equation}
     \sigma_{\rm inel}/\sigma_{\rm tot}= 1 - {\rm Im}a_0\approx 0.83.
\end{equation}
A large portion of inelasticity presumably goes into multibody channels.

What does this imply in the two-body $B$ decays? 
For the factorization-allowed 
processes, the intermediate states would be similar to those which 
appear in elastic meson scattering. Among two-body intermediate states,
those connected to the final state by the Pomeron exchange are the
most important. For the
factorization-forbidden processes in which a quark-pair annihilation 
must occur, the intermediate states of the Pomeron-exchange type are 
forbidden. 
Since the composition of the intermediate states are completely different, 
the factorization-allowed and forbidden amplitudes generally have quite 
different FSI phases even for the same isospin state of, say, $K\pi$.

\section{Two-body intermediate approximation}

In computing the FSI phases, one makes the two-body approximation 
for the intermediate states.\cite{LD} It is a dubious
approximation when inelasticity of scattering is high. 
Getting a number even in this approximation is a daunting task, if one 
tries to do it right.
  
One starts with the unitarity sum of the absorptive part of decay 
amplitude for $B\rightarrow n$ (two-body) and keep only the
two-body intermediate states: 
\begin{equation}
  {\rm Im}M_n = \sum_{n'} a_0(s)_{nn'} M^*_{n'},
      \;\;\; (s=m_B^2; n' = \mbox{\rm two-body}) \label{abs}
\end{equation}
where $a_0(s)_{nn'} (=a_0(s)_{n'n})$ is the $J=0$ 
amplitude for $n\rightarrow n'$. Let us consider only the
factorization-allowed processes. The intermediate states 
connected to $n$ by Pomeron exchange dominate in the sum.
Therefor all $a_0(s)_{nn'}$ 
are purely imaginary. Then we can easily show that $M_n$ are 
also all purely imaginary. This is a special case where the FSI 
phases do not depend on $\cal O$$_i$. 

Phase differences of $M_n$ arise from the non-Pomeron exchanges, $\rho$,
$a_2,\cdots$. Computing 
the non-Pomeron contributions in Eq.(\ref{abs}) is nearly impossible:
First of all, the individual terms in the unitarity sum 
are not real. Only the sum total is. One has to know the phases of $M_{n'}$.
This is circular. To cut the circle, one might approximate
$M_{n'}$ with the factorization amplitudes. That is the Born approximation
valid only for small phases. After obtaining ${\rm Im}M_n$, one has to
compute ${\rm Re}M_n$ by a dispersion integral of ${\rm Im} M_n$ 
over all energies. Even if one could compute for all two-body intermediate 
states, one has no idea as to how much the multibody states modify the
answer.
 
\section{Random approximation}

Computing the FSI phases is difficult because there are so many intermediate
states that we know very little about. Can we take advantage of the presence 
of many states? A statistical approach or a random approximation 
comes to our mind.  Such an attempt was actually made with a limited 
success.\cite{SW}

Start with the unitarity sum, this time including all 
intermediate states:
\begin{equation}
 {\rm Im}M_n = \sum_{n'} a_0(s)_{nn'} M^*_{n'},\;\;(s=m_B^2; n'={\rm all}). 
\end{equation}
Each term in the sum is 
a product of the pure strong-interaction amplitude and the weak decay 
amplitude. Let us postulate that the sign of their product is random from 
one intermediate state to another when there are very many states. 
This is a dynamical hypothesis. It is not true if SD dominates. Under 
this postulate one can compute the statistically likely values of 
the FSI phases as their standard deviations from zero. The likely value  
$\overline{\delta}$ of the FSI phase
depends on the ``favoredness'' parameter $\rho$. It
parametrizes how much favored or disfavored the decay modes are for a given
decay operator: $\rho <1$ for the favored processes such as
the factorization-allowed and color-favored modes, while $\rho<1$ for the
unfavored processes such as factorization-forbidden and/or color-suppressed    
modes. The relation between $\overline{\delta}_n$ and $\rho$ is given by
\begin{equation}
      \tan^2\overline{\delta}_n 
 = \frac{\tau^2(\rho^2-\tau^2)}{1-\rho^2\tau^2},\;\;\;
                        (\tau<\rho<\frac{1}{\tau})  \label{phase}
\end{equation}
where $\tau^2$ is a number determined by elasticity of the final-state
rescattering ($\tau^2 \approx 0.20$ for $K\pi$). The likely phase 
$|\overline{\delta}_n|$ of $M_n$ is plotted in Fig.1. It is small
for the favored modes but can be large for unfavored modes. 
For the ``typical processes'' of $\rho = 1$, the value is $\simeq 20^{\circ}$.
It means that the FSI phases of the typical processes are most likely
between $-20^{\circ}$ and $20^{\circ}$.
\vskip -1cm
\begin{figure}[h]
\epsfig{file=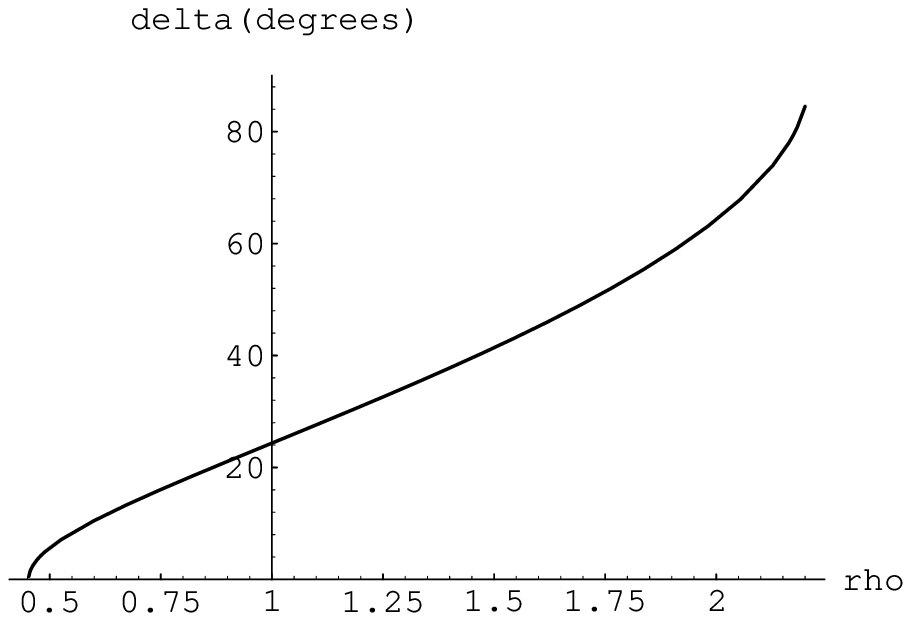,width=6cm,height=3.5cm}
\vskip -1.5cm \hskip 7cm
\epsfig{file=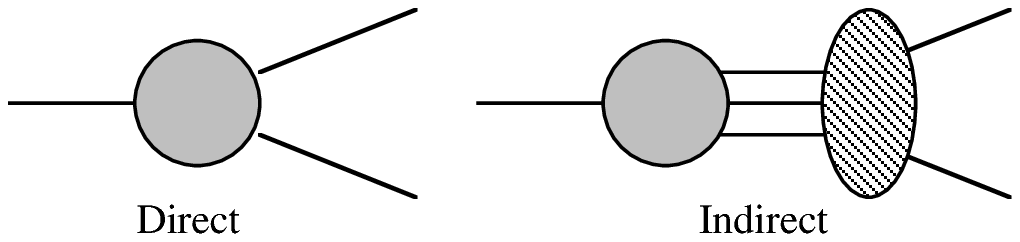,width=4.5cm,height=1.5cm}
\caption{The likely values of the FSI phase {\it vs}
the favoredness parameter $\rho$.}
\label{Fig.1}
\end{figure}
This trend of $\overline{\delta}$ can be understood as follows: For the 
favored modes, the direct decay dominates over the indirect decay going 
through on-shell intermediate states.  Since the 
latter is the origin of the FSI phase, the FSI phase is small. 
For the disfavored modes, the indirect decays dominate since the direct 
decay is suppressed. The random approximation quantifies this
feature in terms of the favoredness parameter $\rho$, given elasticity
of FSI. A similar qualitative reasoning was made by Rosner \cite{Ros}
in the quark picture.  

The limitation of the random approximation is that we cannot 
predict values of individual phases, but only their likely values.

\section{Factorization and enhancement by FSI}
\subsection{Enhancement by FSI}
    The FSI not only generates a strong phase but also changes a
magnitude of amplitude. Fermi \cite{Fermi} argued in potential scattering that 
the FSI should modify the amplitude by the wave function at origin
$\varphi(0,k)$  of final two particles:
\begin{equation}
      M \rightarrow \varphi(0,k)M.
\end{equation}
For slow final particles, the attractive force enhances 
the decay by pulling them close together while the repulsive
force suppresses by pushing them away. Since $\varphi(0,k)$ 
is equal to the Jost function $1/f(k)^*$, it also proves
the phase theorem.  

 This theory can be extended to relativistic rescattering.
In relativistic particle physics, we use the phase-amplitude 
dispersion relation generically referred to
as the Omn\`{e}s-Mushkelishvili representation \cite{Omnes}:
\begin{equation}
   M(s+i\epsilon) = P(s)\exp\biggl(\frac{1}{\pi}\int^{\infty}_{s_{\rm th}}
     \frac{\delta(s')}{s'-s-i\epsilon}ds'\biggr),    \label{OM}
\end{equation}
where $s$ is the $B$ mass squared taken as a dispersion variable, and
$P(s)$ is a polynomial arising from zeros of $M(s)$. $P(s)$ is 
related to the factorization amplitude since it contains no FSI.  
Therefore the FSI enhancement factor of the factorization amplitude 
is given by\footnote{
For $s'>(m_{B^*}+m_{\pi})^2$, $\delta(s')$ contains the initial-state 
interaction or dissociation, part of which enters the form factor of 
the factorization amplitude. It is understood that the initial-state 
interaction contributing only to the form factor has been removed from 
$\delta(s')$ of Eq.(\ref{enhance}).}
\begin{equation}
  E = \exp\biggl(\frac{{\cal P}}{\pi}\int^{\infty}_{s_{\rm th}}
    \frac{\delta(s')}{s'-m_B^2}ds'\biggr).  \label{enhance}
\end{equation}
This form of enhancement was actually proposed and studied for
elastic rescattering at low energies in the late 1950's.\cite {Jacob}
The enhancement $E$ includes the FSI of both SD and LD.
While the low energy portion of the phase integral is purely LD for
an obvious reason, the high energy portion is not entirely SD.

\subsection{Numerical exercise}
I first illustrate the enhancement factor with the decay 
$K\rightarrow \pi\pi$. Since the FSI phase is equal to the scattering 
phase shift below the inelastic threshold, we can evaluate the low-energy
portion of the phase integral up to $\simeq 1$ GeV with the $\pi\pi$ phase
shifts. The result is $E = 2.2$ and $0.8$ for $I=0$ and $I=2$
channels, respectively, which amounts to the relative enhancement of 
$M_{\Delta I=1/2}/M_{\Delta I=3/2} =2.9$. This is a LD
enhancement independent of ${\cal O}$$_i$. Above the inelastic threshold,
$\delta(s')$ is not longer equal to the phase shift and depends on decay
operators.  Therefore the integral above 
1 GeV provides an operator-dependent enhancement
similar to the SD enhancement of RG, but not identical to it.

Let us turn to the $B$ decay. 

In the two-body intermediate approximation, the FSI phase approaches 
$\pm 90^{\circ}$ at high energies. Such an amplitude potentially acquires 
a dangerously large enhancement or suppression since the phase integral can be 
large. If the factorization gives a right ballpark value for a decay amplitude,
$\delta(s')$ must not stay too large at high energies. In the random 
approximation, the sign of $\delta(s')$ may fluctuate with $s'$.  
In this case, the effect of enhancement and suppression would be much smaller.

Here I present a numerical estimate for $B\rightarrow K\pi 
(\pi\pi)$  with one model FSI phase motivated by experiment. 
Knowing that the phases are small for $D\rightarrow\overline{K}\pi$
but possibly small for $B\rightarrow K\pi$, I choose the FSI
phase as shown in Fig.2:
$\delta(s')$ rises to large values ($\sim 90^{\circ}$) around 
2 GeV and falls linearly to zero at $E_{\rm max}= O(m_B)$.\footnote{To
be precise, this $\delta(s')$ includes part of the initial-state 
interaction above $s'=(m_{B^*}+m_{\pi})^2$. 
See the footnote in the preceding page.} 
\begin{figure}[h]
\epsfig{file=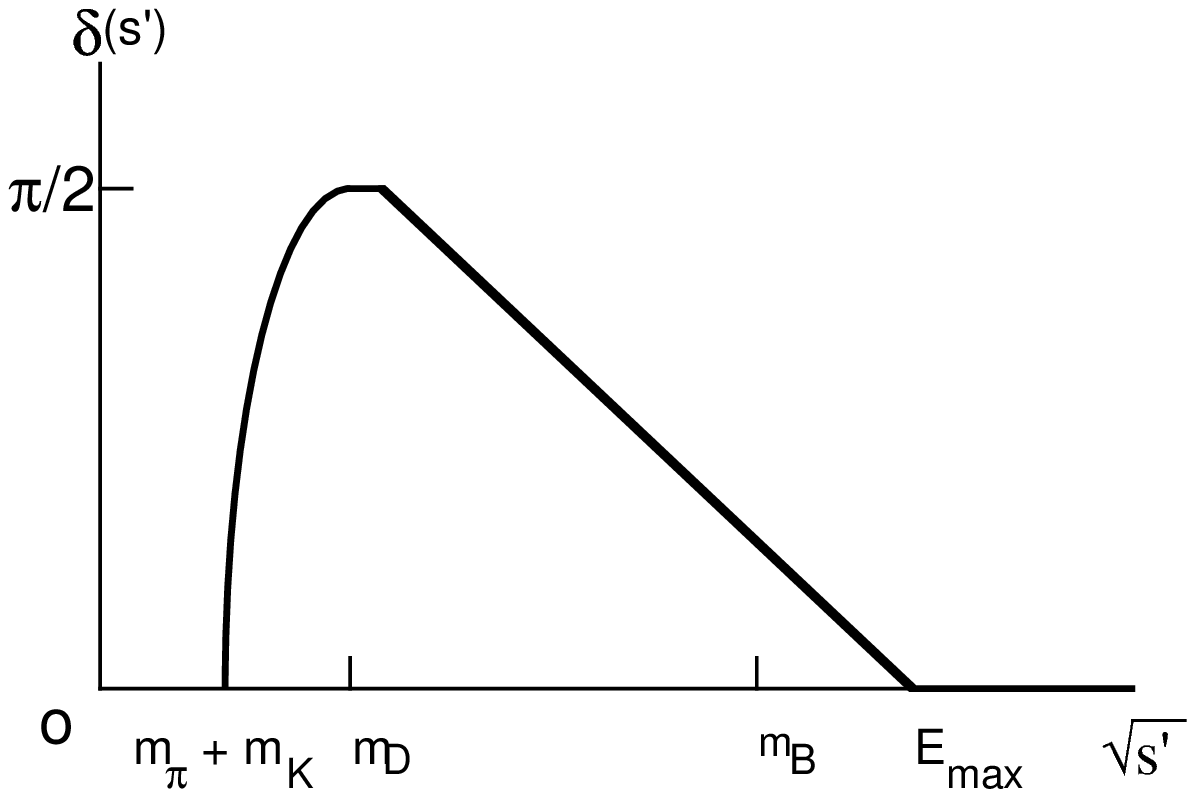, width=5cm,height=4.5cm}
\vskip -6cm \hskip 5cm
\epsfig{file=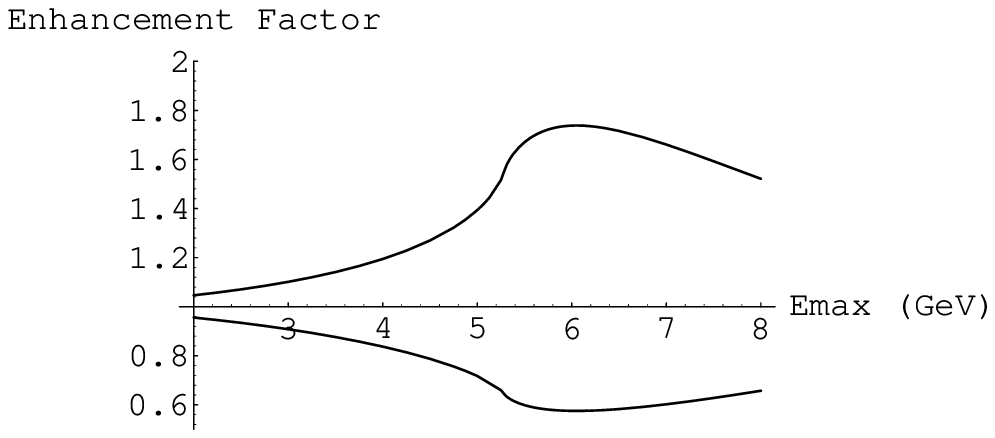,width=6.5cm,height=4.5cm}
\caption{The model FSI phase and the enhancement (suppression) factor 
$E$ vs $E_{max}$. The lower curve ($E<1$) for $\delta(s')>0$ shown here,
and the upper curve for $\delta(s')\rightarrow -\delta(s')$.}
\label{Fig.2}
\end{figure}
This $\delta(s')$ suppresses the decay amplitude since the support 
of the phase integral is mostly below $m_B$ where  
$1/(s'-m_B^2)$ is negative. The suppression factor is given by the lower 
curve in the right of Fig.2. It can easily a 10$\sim$40\% correction
to the factorization as $E_{\rm max}$ varies. 

We learn two lessons from this exercise.
Unlike the phase, the magnitude of amplitude receives the FSI
effect from all energies. Even if the FSI phase happens to be 
zero at $m_B$, the FSI can generate a substantial correction to 
the factorization amplitude. Next, the low-energy (LD) contribution 
is power-suppressed in $1/m_B^2$, but not necessarily like 
$\Lambda_{\rm QCD}^2/m_B^2$. It may be more like $m^2_D/m_B^2$
or $m_{J/\psi}^2/m_B^2$ ($= 15\sim 30 \%$).  

\section{Conclusion}

   If LD physics is important in two-body $B$ decays, computing individual FSI
phases is nearly an impossible proposition. What we can predict at best
is their statistically likely values. 

The phase and the magnitude of amplitude are related by analyticity. 
Multiplying a large FSI phase on a factorization amplitude
is very likely inconsistent with analyticity.
If an amplitude is significantly enhanced or suppressed
relative to its factorization value, it indicates that the FSI phase 
is large and LD physics is important.  When we attempt to extract the CP 
phases from experiment, we should be apprehensive about the level of 
accuracy of factorization 
amplitudes. We must not use calculated values of FSI phases in analysis.

\section*{Acknowledgment}
 This work was supported in part by the Director, Office of Science,
Office of High Energy and Nuclear Physics, Division of High Energy Physics,
of the U.S. Department of Energy under contract
DE-AC03-76SF00098 and in part by National Science Foundation 
under grant PHY-95-14797.

\end{document}